\documentclass[preprint,5p,twocolumn]{elsarticle}

\usepackage[utf8]{inputenc}
\usepackage[T1]{fontenc}
\usepackage{textcomp}
\usepackage{lineno}
\usepackage[colorlinks=true]{hyperref}
\usepackage{siunitx}
\usepackage{amsmath}
\usepackage{xspace}
\usepackage{xcolor}

\modulolinenumbers[5]

\journal{Nuclear Instruments and Methods in Physics Research~A}

\hypersetup{
  pdfstartview={XYZ null null 1},
  pdfauthor={A. Hollering, N. Rebrova, C. Klauser, Th. Lauer, B. Märkisch, U. Schmidt},
  pdftitle={A non-depolarizing CuTi neutron supermirror guide for PERC}
}

\newcommand{\Nitro}{N\textsubscript{2}\xspace}
\newcommand{\TiNx}{$\mathrm{TiN}_x$\xspace}

\begin{document}
\begin{frontmatter}

\title{A non-depolarizing CuTi neutron supermirror guide for PERC}

\author[1,2]{A. Hollering}
\ead{alexander.hollering@tum.de}
\author[3]{N. Rebrova}
\ead{rebrova@jinr.ru}
\author[4]{C. Klauser}
\ead{Christine.Klauser@psi.ch}
\author[5,2]{Th. Lauer\fnref{fn1}}
\author[1]{B. Märkisch\corref{corresponding}}
\ead{maerkisch@ph.tum.de}
\author[3]{U. Schmidt}
\ead{ulrich.schmidt@physi.uni-heidelberg.de}

\cortext[corresponding]{Corresponding author}
\fntext[fn1]{Present address: MagTec GmbH, Bahnstraße 73, 67158 Ellerstadt}

\address[1]{Physik-Department ENE, Technische Universität München, James-Franck-Str.1, D-85748 Garching, Germany}
\address[2]{Forschungsneutronenquelle Heinz Maier-Leibnitz, Lichtenbergstr. 1, D-85748 Garching, Germany}
\address[3]{Physikalisches Institut, Im Neuenheimer Feld 226, Universität Heidelberg, 69120 Heidelberg, Germany}
\address[4]{Institut Laue-Langevin, 71 avenue des Martyrs, F-38042 Grenoble, France}
\address[5]{Movatec GmbH, Erfurter Straße 23, D-85386 Eching, Germany}

\begin{abstract}
Neutron guides are used to transport slow neutrons from sources to experiments. Conventional neutron supermirror guides use alternating thin layers based on nickel and titanium. Due to the magnetic properties of nickel, their neutron reflection properties are spin-dependent, in particular when exposed to high magnetic fields. Motivated by the requirements of precision experiments on neutron beta decay, we present novel supermirrors based on copper and titanium, which preserve the neutron beam polarization. These show excellent reflectivity and prove to be very stable even when exposed to high temperatures.
\end{abstract}


\end{frontmatter}


\section{Introduction}

Neutron guides are long glass channels whose inner surfaces are coated with a reflecting layer for angles of incidence up to a critical angle. They allow to transport cold and thermal neutrons away from a reactor or spallation source to experimental sites as far as $\SI{200}{m}$ away from the source~\cite{HolmDahlin19, Andersen20}.
\emph{Supermirrors}, first devised by Mezei and Dagleish \cite{Mezei77}, consist of a large number of thin double layers of typically Ni/Ti and varying in thickness such that the Bragg condition for neutrons with grazing incidence is fulfilled for a broad band of neutron de~Broglie wavelengths. They hence increase the maximum angle of reflection at a given wavelength beyond the critical angle of total reflection of the best element Nickel by some factor $m$.

Within the neutron decay experiment PERC currently under construction at the MEPHISTO \cite{MEPHISTO} beam line of the Heinz Maier-Leibnitz Zentrum (MLZ), Garching, Germany, an $\SI{8}{\metre}$ long guide inside an approximately $\SI{12}{\metre}$ long magnet system serves as decay volume \cite{Dubbers08,Wang19}, with the guide surface largely in direction of the magnetic field. PERC will measure correlation coefficients in neutron $\beta$-decay in order to precisely determine the nucleon weak axial-coupling $g_\mathrm{A} = \lambda  g_\mathrm{V}$, the first element of the quark-mixing Cabibbo-Kobayashi-Maskawa matrix $V_{ud}$ and to search for physics beyond the Standard Model of particle physics like novel scalar and tensor couplings. For a recent review of the field see \cite{Dubbers21}.

In a guide, neutrons are reflected on the walls due to certain Bragg conditions that are defined by the layer thicknesses and material composition of a multilayer supermirror system. In the PERC experiment, the neutron spin of a highly polarized neutron beam has to be preserved on the level of $10^{-4}$ per bounce in order to match the precision requirements, see \cite{Dubbers08}. Supermirror coatings made of layers of ferromagnetic Ni and layers of paramagnetic Ti can depolarize the polarized neutron beam due to magnetic scattering if the magnetization has a component perpendicular to the scattering vector inside the coating \cite{Padiyath04}. Nickel with added molybdenum is commonly used for non-depolarising neutron guides \cite{Padiyath04,Schebetov99,KovacsMezei08}. Single layers of NiMo and NiP alloys are used for non-depolarizing guides for ultra-cold neutrons (UCNs), as well as a number of other materials, see \cite{Bondar17, Tang16} and references therein.

In this paper, we present a novel neutron supermirror system made of Cu/Ti layers, the optimization procedure and its properties. A coating made of diamagnetic Cu and weakly paramagnetic Ti should be nearly non-magnetic even in strong magnetic fields of $\approx \SI{2}{T}$ and can hence maintain the neutron polarization on the highest level.

In order to optimize the new supermirrors, we make use of different characterization techniques. We present the experimental results of various Cu/Ti single layers, multilayers of alternating bilayers of constant thickness, and neutron supermirrors produced by pulsed DC and RF magnetron sputtering.
We studied the effects of pure nitrogen inside the Ti with X-ray reflection (XRR) and X-ray diffraction (XRD). The thermal stability was evaluated by neutron reflectivity measurements before and after heat treatment. The model developed for the roughness growth inside a supermirror structure agrees well with the experimental data. The material composition of the layers was investigated using elastic recoil detection (ERD) \cite{Bergmaier95} at the Q3D instrument \cite{Dollinger04} of the Maier-Leibnitz Laboratory (MLL), Garching. Depolarization measurements of early supermirrors were performed at the PF1b beam line \cite{Abele06} of the Institut Laue-Langevin (ILL), Grenoble, France, using an opaque test bench with polarized \textsuperscript{3}He spin-filter cells \cite{Klauser13a,Klauser16}.

\section{Experimental details}

The substrate material has max. dimensions of $200 \times \SI{300}{\milli\metre^2}$ and the max. capacity of the sputtering machine is $300 \times \SI{1200}{\milli\metre^2}$.
The Cu films were deposited by RF magnetron sputtering with a frequency of $\SI{13.56}{\mega\hertz}$. The Ti films were deposited using pulsed DC magnetron sputtering with a frequency of $\SI{50}{\kilo\hertz}$ and an additional substrate bias voltage. The base pressure of the system was $\SI{8e-7}{\milli\bar}$ and the pressures during sputtering were $\SI{5.4e-3}{\milli\bar}$ for Ti and $\SI{1.9e-3}{\milli\bar}$ for Cu. The sputtering power for both materials was set to $\SI{500}{\watt}$. The distance between substrate and sputtering head was $\SI{150}{mm}$. During the Ti sputter process, \Nitro gas was admixed with the Ar sputter gas.

The samples consisting of \TiNx single layers (with $x$ up to $\approx 30\,\%$)  and Cu/\TiNx multilayers and neutron supermirrors were deposited on float glass. Before coating, the glass mirrors were first cleaned in an ultra-sonic bath using alkaline mucasol cleaner and then processed further using isopropanol. In the sputtering chamber, they were cleaned in-situ using an Ar ion beam. The \TiNx single layers have thicknesses of $\SI{32}{\nano\metre}$, $\SI{165}{\nano\metre}$, and $\SI{300}{\nano\metre}$. The multilayers consist of $40$ bilayers of equal thickness of $\SI{6.6}{\nano\metre}$ for \TiNx and $\SI{6.9}{\nano\metre}$ for Cu. The neutron supermirrors consist of $95$ bilayers Cu/\TiNx with varying thickness.

The density, thickness and surface roughness of each film were determined by XRR measurements in $\Theta/2\Theta$ mode using Cu$\text{K}_{\alpha}$ radiation ($\SI{1.54}{\angstrom}$) on a Bruker D8 diffractometer. The lattice parameters were determined by XRD measurements on the same device in $\Theta/2\Theta$ mode and in gracing incidence mode (GIXRD).

The neutron reflectivity (NR) curves of the multilayers and supermirrors were measured at the instrument \mbox{TREFF} at FRM~II \cite{TREFF}.
The NR investigations were performed at a fixed neutron wavelength of $\SI{4.8}{\angstrom}$ in $\Theta/2\Theta$ mode. The resistance to heat treatment of the neutron supermirrors was also verified by NR at \mbox{TREFF}.

\section{Results and discussion}

\subsection{Effects of impurities on single- and multilayers}

A previous attempt at a Cu/Ti based supermirror with 20 bilayers is documented in Ref.~\cite{Pleshanov94}. The quality of the data did not allow conclusions on the effect of the high mobility of Cu, which we find to be significant as shown below.
The \TiNx layers can form an effective interdiffusion barrier \cite{Wang90,Olowolafe92,Kwak99,Kim00}. Using air as reactive sputtering gas as is commonly done for Ni layers \cite{SenthilKumar97} could lead to ferromagnetic clusters caused by copper oxides \cite{Radjehi18}. We consequently investigated the effects of a pure nitrogen admixture on the microstructure of the Ti layers by using XRD on samples with a nominal thickness of $\SI{300}{\nano\metre}$ deposited on float glass.

\begin{figure}[tb]
\centering
\includegraphics[width=.45\textwidth]{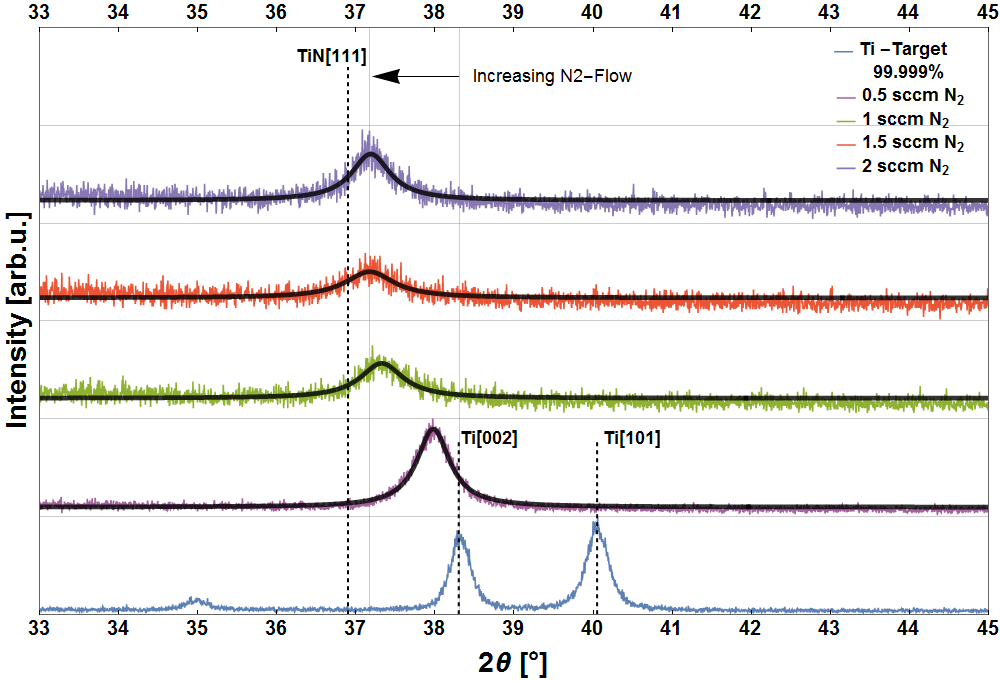}
\caption{XRD on single layers of Ti and \TiNx with varying nitrogen flow during deposition. The black lines correspond to the fits of peak positions listed in Tab.~\ref{tab:XRDfit}. }
\label{fig:XRD}
\end{figure}

The addition of nitrogen to the Ar sputtering gas leads to a dilatation of the Ti lattice. The pure Ti target material shows two preferential planes, namely [002] and [101]. The Ti layer grown under pure Ar atmosphere shows the [002] plane, which gradually changes to the [111] plane of the TiN compound  \cite{Khojier13} with increasing nitrogen content. The [101] plane of pure Ti disappears. Generally, the reactively deposited Ti layers tend to grow more in amorphous structure. Fig.~\ref{fig:XRD} shows the results of the XRD measurements on samples with different admixtures of \Nitro.

\begin{table}[tb]
	\centering
  \begin{tabular}{ c | c c }
    \text{nitrogen flow [sccm]} &  \text{peak position [\textdegree]} & peak width [\textdegree]    \\
    \hline
    $0  $  &  $\SI{38.321+-0.001}{}$ &  $\SI{0.136+-0.001}{}$  \\
    $0.5$  &  $\SI{37.978+-0.003}{}$ &  $\SI{0.251+-0.005}{}$  \\
    $1  $  &  $\SI{37.326+-0.020}{}$ &  $\SI{0.33+-0.03}{}$  \\
    $1.5$  &  $\SI{37.173+-0.031}{}$ &  $\SI{0.38+-0.04}{}$  \\
    $2  $  &  $\SI{37.187+-0.015}{}$ &  $\SI{0.30+-0.02}{}$  \\
    \end{tabular}
\caption{\label{tab:XRDfit}Peak positions obtained by fitting the XRD data from TiN$_x$ single layers shown in Fig.~\ref{fig:XRD}. The nitrogen flow during sputtering is given in \emph{standard cubic centimeters per minute} (sccm).}
\end{table}

We expect that the observed change in crystal structure and the nitrogen inside the Ti limit the interdiffusion for the multilayer structure and reduce roughness growth from layer to layer. As a downside, an increasing \Nitro concentration also lowers the neutron optical contrast between the layers.

\begin{figure}[tb]
\centering
\includegraphics[width=.45\textwidth]{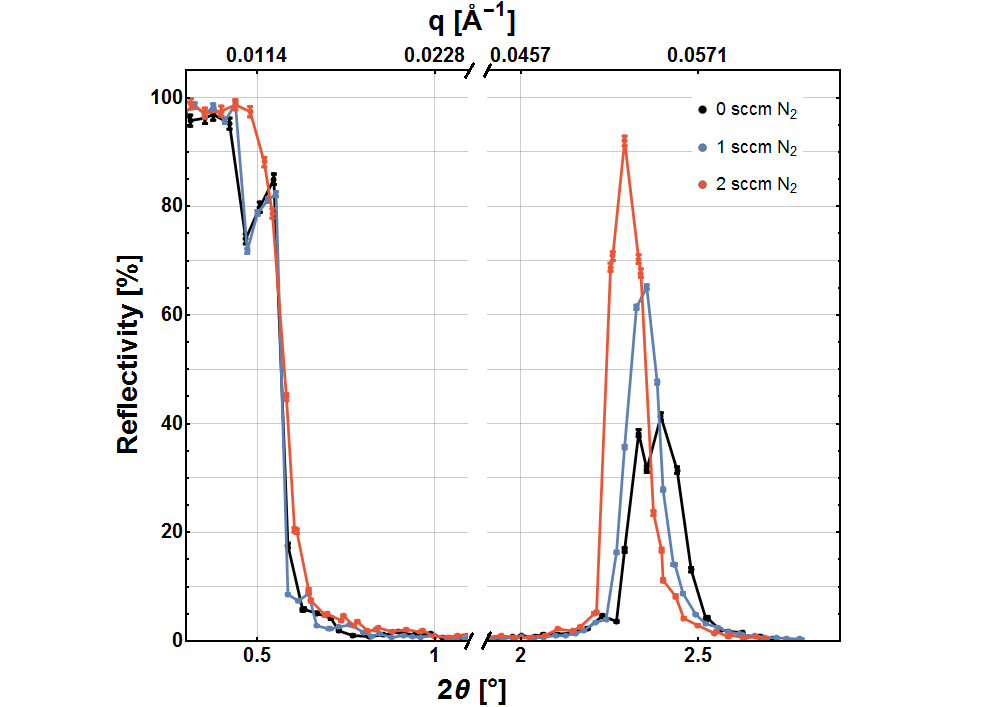}
\caption{Neutron reflectivity curves of three multilayers with varying \Nitro gas flow during sputtering. To help distinguish the peaks, data points are connected with colored lines. We observe total reflection from the top layer as well as a single Bragg peak from the multilayer with all bilayers of equal thickness. According to our modeling, the behavior at the edge can qualitatively be attributed to additional interference due to the inter-layer diffusion.}
\label{fig:Monochromators}
\end{figure}

Fig.~\ref{fig:Monochromators} shows three NR measurements of multilayers deposited with different nitrogen admixtures. As the mirrors each consist of 40 bilayers with equal layer thicknesses, we obtain constructive interference at a particular Bragg angle of reflection, as well as total reflection from the top layer below a critical angle.

The interpretation of such NR measurements is non-trivial. The roughness of the layers and interdiffusion of the different layer materials have very similar impact on the neutron reflectivity. It is hence not possible to distinguish between them from NR data. The high neutron reflectivity and the smallest peak width of the mirror sputtered with $2\,\mathrm{sccm}$ \Nitro indicate that both, interlayer roughness and interdiffusion, are the lowest for this sample.  With lower \Nitro concentration, the peak width increases and the peak position shifts to larger angle of reflection. If the  interface roughness increases, the proportion of diffuse scattered neutrons increases, which broadens the reflected peak. Assuming that two layers diffuse into one another, the effective thickness of the layers \emph{`seen'} by the incoming neutrons gets smaller. The thinner the layers are, the larger is the angle of Bragg reflection.

\subsection{Single surface roughness}

\begin{figure}[tb]
\centering
\includegraphics[width=.45\textwidth]{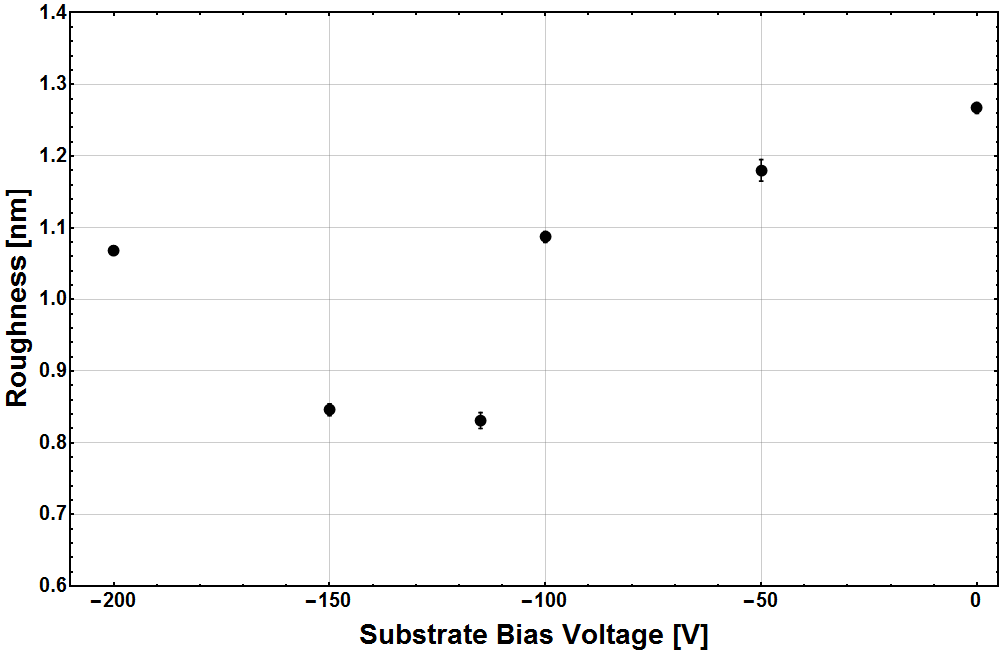}
\caption[]{The surface roughness of \TiNx single layers for a \Nitro gas flow of $\SI{2}{sccm}$ and with a thickness of $\SI{32}{\nano\metre}$ varies with the applied bias voltage and we observe a minimum at $\approx-115 \ldots\SI{150}{\volt}$. The roughness of the glass substrate is $\approx \SI{0.6}{nm}$.}
\label{fig:RoughnessBias}
\end{figure}

Based on the above results, we conclude that sputtering Ti reactively with $2$\,sccm \Nitro would be the most promising option to obtain a Cu/\TiNx neutron supermirror with highest possible reflectivity. We investigate the influence of a substrate bias voltage on the surface roughness of single \TiNx layers. Fig.~\ref{fig:RoughnessBias} shows the roughness of six \TiNx single layers deposited on float glass with a thickness of $\SI{32}{\nano\metre}$ each, as calculated with our model, see section \ref{sec:RoughnessModel}.

The surface roughness shows a clear dependency of the substrate bias voltage, which can be explained as follows: By applying a negative bias voltage to the substrate, some fraction of the ionized Ar atoms are accelerated onto the substrate surface. The surface roughness of the layer is the smallest between $\SI{-115}{\volt}$ and $\SI{-150}{\volt}$, when the ions have enough energy to remove loosely bound atoms to smooth the surface. This can be interpreted as a nanopolishing procedure. For a bias voltage between $\SI{-100}{\volt}$ and $\SI{0}{\volt}$, the surface polishing effect decreases with the smaller energy of the Ar ions. For a voltage below $\SI{-150}{\volt}$, the ions have enough energy to kick out Ti atoms from the deposited layers. This effect is called re-sputtering and causes the layer roughness to increase \cite{Logothetidis96}.

\begin{figure}[tb]
\centering
\includegraphics[width=.45\textwidth]{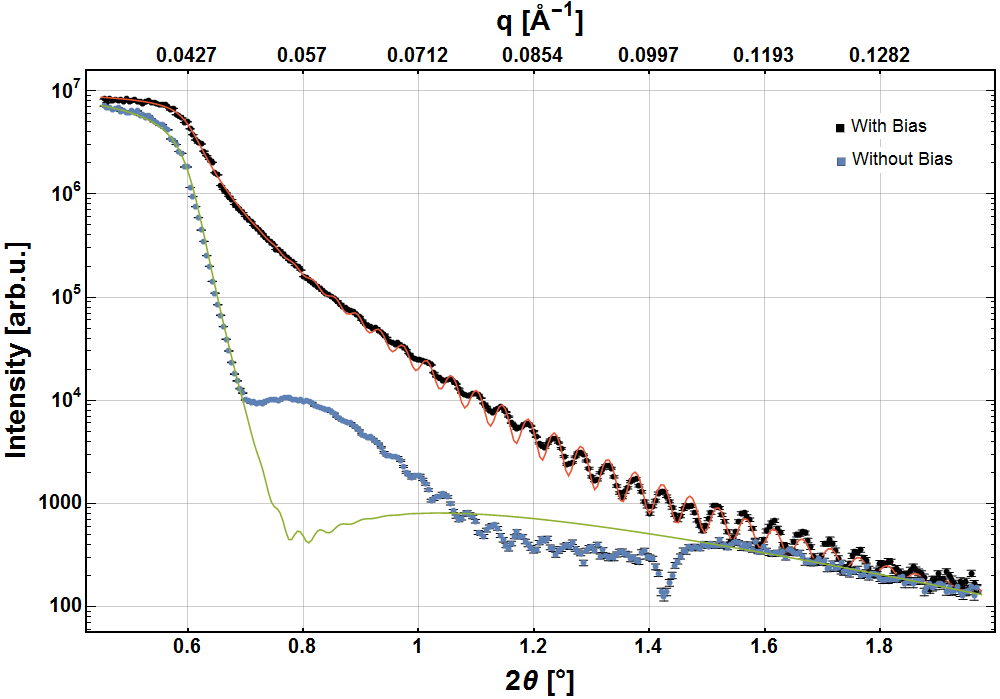}
\caption{XRR measurement of $\SI{165}{\nano\metre}$ thick \TiNx layers deposited with a \Nitro gas flow of $\SI{2}{sccm}$ and with or without substrate bias voltage. The red curve shows the fit result, with a surface roughness of only $\SI{2.15+-0.03}{\nano\metre}$. The green curve is not a fit, but a calculation assuming a roughness of $\approx\SI{6.5}{\nano\metre}$.
}
\label{fig:RoughnessMono}
\end{figure}

The effect of applying a substrate bias voltage of $\SI{-150}{\volt}$ is confirmed by the XRR measurement results shown in Fig.~\ref{fig:RoughnessMono}. A $\SI{165}{\nano\metre}$ thick \TiNx layer, deposited without bias voltage is compared to a $\SI{167}{\nano\metre}$ layer deposited with additional bias voltage of $\SI{-150}{\volt}$. For the sample deposited without bias voltage, the green curve in the figure, which is calculated according to our model, indicates a surface roughness of $\approx\SI{6.5}{\nano\metre}$. The difference between the green theoretical curve and the measurement is presumably coming from an $\approx\SI{7}{\nano\metre}$ thick \TiNx interlayer formed by oxidation in the near surface region. For the sample sputtered with bias voltage, a surface roughness of $\SI{2.15+-0.03}{\nano\metre}$ was determined (red curve). Taking the roughness of the float glass substrate of $\SI{0.6}{\nano\metre}$ into account, and assuming a linear growth of roughness with thickness, this corresponds to a roughness growth rate of only $\SI{0.09}{\angstrom\per\nano\metre}$.

\subsection{Neutron supermirror}

\begin{figure}[tb]
\centering
\includegraphics[width=.45\textwidth]{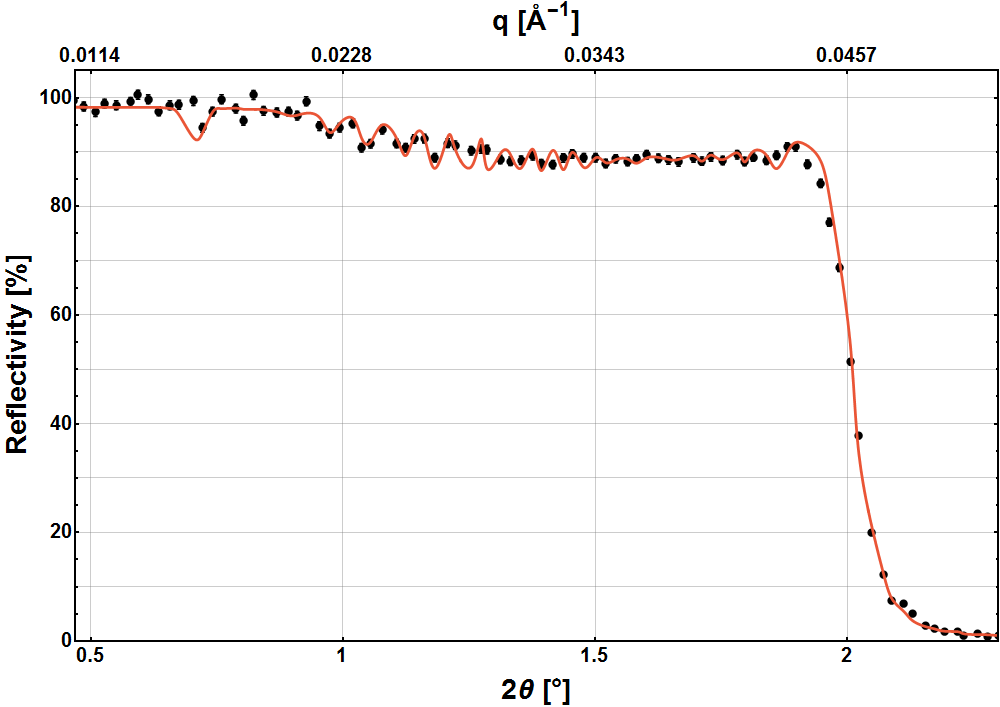}
\caption[]{Neutron reflectivity measurement of a Cu/\TiNx $\text{m}\,=\,2$ neutron supermirror with reflectivity of above $90\,\%$. The red curve is a fit using our roughness growth model in Eq.~\eqref{eq:RoughnessModel}. The reduced $\chi^2$-value of the fit is $6.8$.}
\label{fig:Supermirror}
\end{figure}

Taking into account our investigations on single- and multilayers, as well as improving the sputtering conditions, we produced Cu/\TiNx neutron supermirrors with a high  reflectivity of $> 90$\,\% at the critical angle of reflection in the best case. For the sample shown in Fig.~\ref{fig:Supermirror}, this reflectivity is $92\,\%$.

The layer sequence for the mirrors was calculated by the following method: We approximate the layer sequence we obtain according to Ref.~\cite{Mezei77} for each material separately by a 5 point interpolation function. Then we calculate the reflectivity including our roughness growth model and optimize the reflectivity by varying the supporting points of the sequence interpolating function with a stochastic annealing method.
High-purity sputter targets (Cu $99.995\,\%$, Ti $99.999\,\%$) were used for the supermirror coatings to be free of any critical impurity like e.g. iron. The reflectivity measurements in Fig.~\ref{fig:Supermirror} and Fig.~\ref{fig:SupermirrorHeat} were performed in single reflection mode. The critical angle of reflection corresponds to $m=2$, i.e. double the critical angle of natural nickel.

\subsection{Temperature stability}

Stability of the mirrors at elevated temperatures is important for two reasons. First, it indicates that the interdiffusion is indeed reduced by the addition of N to the Ti layers. Additionally, the guide will be located in the $\approx\SI{12}{\metre}$ long warm bore of the PERC magnet. To improve the vacuum conditions there, the bore can be backed at $\SI{100}{\degreeCelsius}$, see Ref.~\cite{Wang19}. But this procedure must not degrade the neutron guide.

\begin{figure}[tb]
\centering
\includegraphics[width=.45\textwidth]{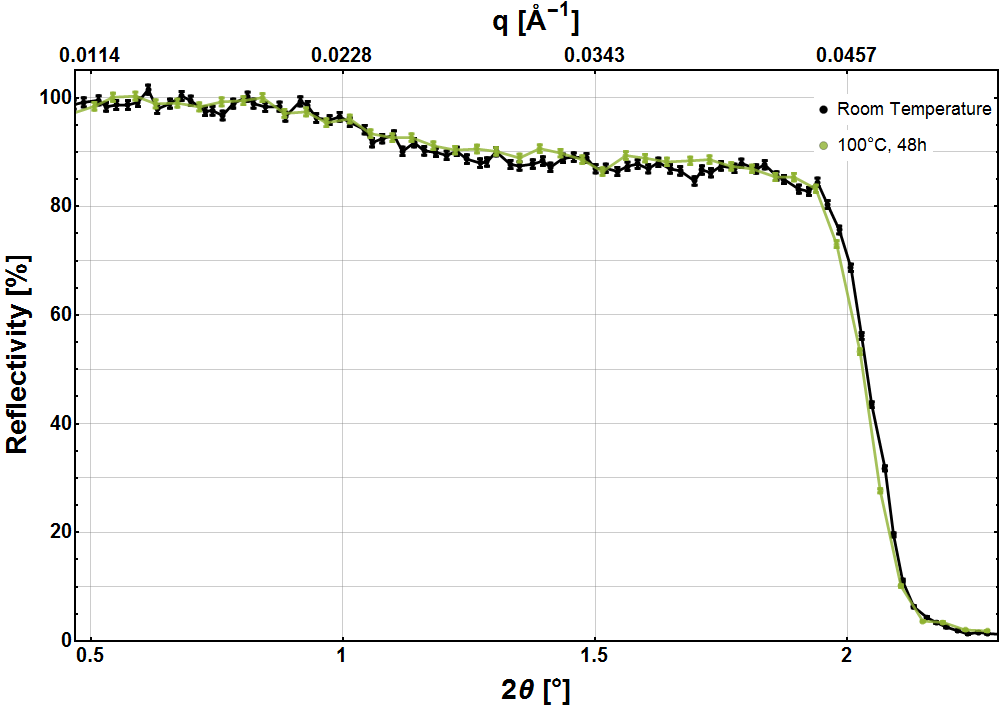}
\caption[]{NR measurement of a Cu/\TiNx $m=2$ supermirror consisting of $95$ bilayers with a reflectivity of $R=85\,\%$ before and after baking for $\SI{48}{\hour}$ at $\SI{100}{\degreeCelsius}$. Measurement points are connected by lines to help distinguish the data sets.}
\label{fig:SupermirrorHeat}
\end{figure}

In Fig.~\ref{fig:SupermirrorHeat} we present the NR measurement of one mirror with slightly lower reflectivity, which was used for thermal stability evaluation. The mirrors were deposited on standard float glass. The supermirror coating consists of $95$ Cu/\TiNx bilayers with varying thickness. The mirror shows an excellent temperature stability. Still after baking for $\SI{48}{\hour}$ at a temperature of $\SI{100}{\degreeCelsius}$, the neutron reflectivity remains practically unchanged.

\begin{figure}[tb]
\centering
\includegraphics[width=.45\textwidth]{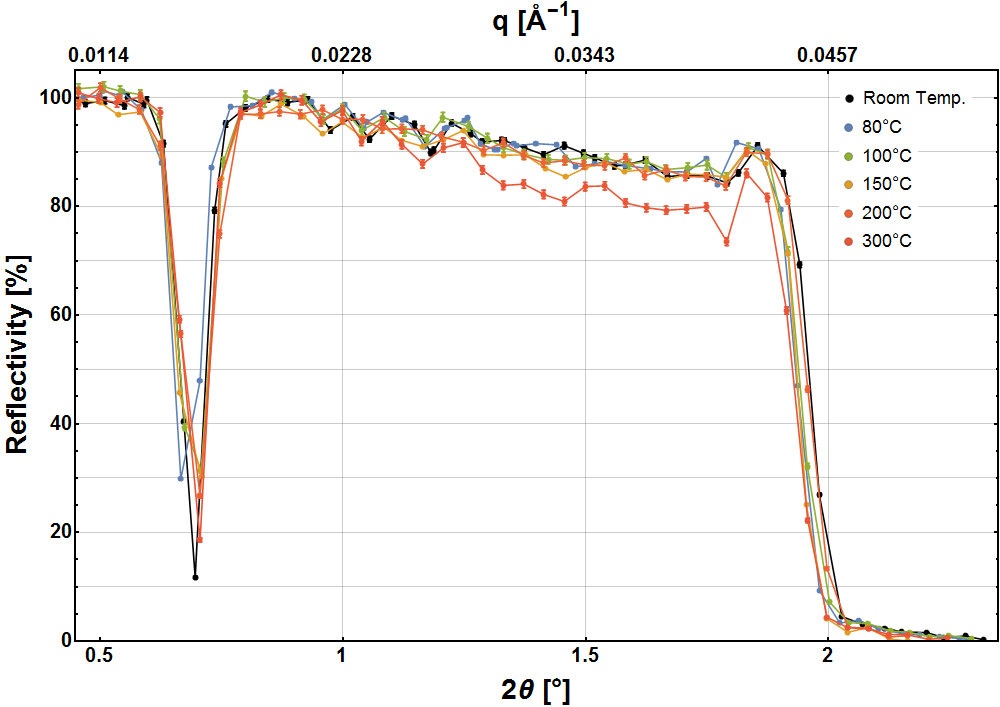}
\caption[]{NR measurement of an early Cu/\TiNx $m=1.95$ supermirror after subsequent baking for $\SI{12}{\hour}$ at $80 \text{ to } \SI{300}{\degreeCelsius}$. Only after baking at \SI{300}{\degreeCelsius} the reflectivity decreases significantly. The dip in reflectivity is due to a missing top-layer. Measurement points are connected by lines to help distinguish the data sets.}
\label{fig:SupermirrorHeatExt}
\end{figure}

Tests at higher temperatures were performed with an earlier sample. This was produced using low-purity targets and unfortunately suffered from a production error, a missing top-layer. The overall performance of this early sample is still worse than the recent result shown in Fig.~\ref{fig:Supermirror}. The sample was subsequently heated for $\SI{12}{\hour}$ at $80, 100, 150, 200 \text{ and } \SI{300}{\degreeCelsius}$, and the neutron reflectivity was measured after each treatment. The results in Fig.~\ref{fig:SupermirrorHeatExt} show that only baking at $\SI{300}{\degreeCelsius}$ has a significant effect.

\subsection{Roughness growth model}
\label{sec:RoughnessModel}

To describe the roughness growth inside the Cu/\TiNx supermirror stack, we need to account for the effect of interdiffusion. The new model was expanded
and integrated into an existing fit routine used for the description of Ni/Ti supermirrors. The roughness of the $i$-th layer is given by the roughness of the substrate $r_\mathrm{sub}$, a constant (effective) roughness due to interdiffusion $r_\mathrm{diff}$ and a term which describes the roughness accretion $r$ with thickness:
\begin{equation}
r_i\,=\, r_\mathrm{sub}\,+\,r_\mathrm{diff}\,+\,r\cdot \left(\frac{d_i}{d_{\text{tot},i}} \right)^p,
\label{eq:RoughnessModel}
\end{equation}
where $d_i$ and $d_{\text{tot},i}$ are the thicknesses of the $i$-th layer and all layers below, respectively, and $p$ is the power of the roughness growth. We include roughness in our reflectivity calculation by folding the step function of the neutron optical potential at the layer borders with a Gaussian function where the standard deviation is given by the roughness except for $r_\mathrm{diff}$. Here, the Gaussian function is replaced by a double-sided decaying exponential which leads to a better approximation of a diffusion profile.

\begin{table}[htbp!]
  \centering
  \begin{tabular}{ c | c }
    $r_\mathrm{sub}$ [nm]   &  $\SI{0.80+-0.02}{}$  \\
    $r$ [nm]                &  $\SI{2.41+-0.09}{}$  \\
    $r_\mathrm{diff}$ [nm]  &  $\SI{0.79+-0.02}{}$  \\
    $p$                     &  $\SI{1.41+-0.06}{}$  \\
    \hline
     red. $\chi^2$ of the fit & $6.8$    \\
    \end{tabular}
\caption{Roughness parameters from a fit to the reflectivity data of the Cu/\TiNx ($\text{m}\,=\,2$, $190$ layers) neutron supermirror shown in Fig.~\ref{fig:Supermirror} as parametrized by Eq.~\eqref{eq:RoughnessModel}. }
		\label{tab:RoughnessFit}
\end{table}

For sputtering without ion plating, the deposited layer follows purely statistical roughness growth with $p = 0.5$, similar to snow flakes accumulating on smooth surfaces. For our Cu/\TiNx supermirrors produced with a bias voltage of $\SI{-150}{\volt}$ we obtain $p\,>\,1$, as the surface is polished in-situ by the incoming accelerated particles. The resulting roughness growth parameters for the red curve shown in Fig.~\ref{fig:Supermirror} are listed in Tab.~\ref{tab:RoughnessFit}.

For a total thickness of the supermirror coating of $\approx\SI{2}{\micro\metre}$, the surface roughness ($r_\text{sub}\,+\,r_\text{diff}\,+\,r$) of the supermirror is only $\SI{4}{\nano\metre}$. We observe the parameter $p > 1$, which indicates that the roughness growth is small in the beginning and increasing with the overall coating thickness. This effect can be explained by the small surface roughness of the substrate, which helps to keep the roughness low from the beginning and as well as by the fact, that the first layers have only a thickness of a few nanometers, where amorphous growth is favored and crystal growth suppressed. This  roughness model might also prove helpful for the understanding of conventional supermirrors with higher $m$-values.

\subsection{Magnetic properties}

The presence of magnetic roughness in layers, as is typical for ferromagnetic materials, causes depolarization of neutrons. We investigate early mirrors produced using a less pure Ti target ($99.5\%$), which is known to contain traces of iron due to the production process. At the PF1b beamline of the ILL, we characterise these samples with $m$-values up to $1.2$ with an in-plane magnetic field of up to $\SI{0.82}{T}$ using an opaque test bench with two \textsuperscript{3}He neutron spin filter cells. The setup is described in Fig.~4 of Ref.~\cite{Klauser16}, where it was used to characterize neutron polarizers. The measurement technique is sensitive to polarization and depolarization effects below the $10^{-4}$ level. With these early mirror samples, we find depolarization of $6(4)\times10^{-5}$. More details on this experiment can be found in \cite{Rebrova14}.

\begin{figure}[tb]
\centering
\includegraphics[width=.45\textwidth]{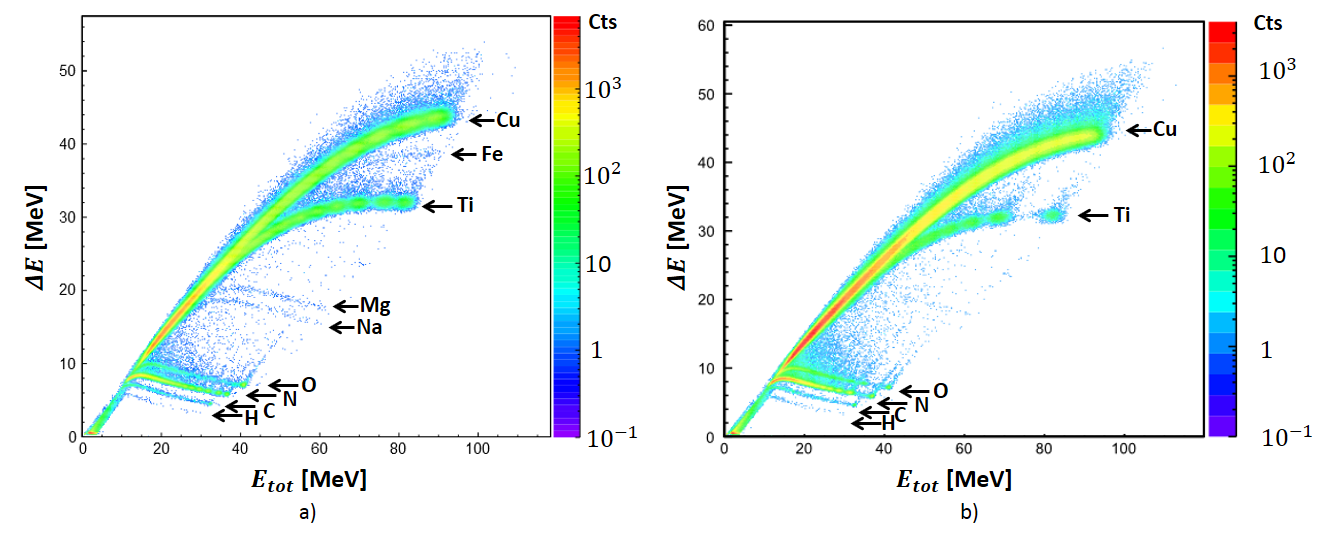}
\caption[]{Element identification with ERD. \textbf{a)} A Cu/Ti supermirror created using targets with purities of $99.95\,\%$ for Cu and $99.6\,\%$ for Ti. Additional impurities like Na, Mg and Fe are detected. \newline \textbf{b)} Measurement of a Cu/Ti supermirror created using targets with purities of $99.995\,\%$ for Cu and $99.999\,\%$ for Ti. Here, no Fe is detectable.}
\label{fig:ERD}
\end{figure}

We note that Ref.~\cite{Bondar17} finds depolarisation at the level of a few times $10^{-5}$ per bounce for UCNs on single NiMo and Cu layers, but the application to our case with cold neutrons and high magnetic fields is not straight-forward.

Given the demanding (de)polarization criteria of the neutron decay measurement with PERC and its high magnetic field of $\SI{1.5}{T}$, and the in total thicker layer sequence of the envisaged $m=2$ supermirror, we switch to highly pure Ti targets for further development. The material composition of the layer sequence was investigated using elastic recoil detection (ERD) using an $\SI{170}{MeV}$ \textsuperscript{127}I ion beam at the Q3D instrument at the MLL.
The impurities Na, Mg and Fe, which were found in the Ti Target with $99.6\,\%$ purity, can be explained by the Kroll production process \cite{Kroll40} and are consistent with the material's technical data sheet. In the Ti targets with $99.999\,\%$ purity, no Fe impurity was detectable, see Fig.~\ref{fig:ERD}.
Such targets are typically produced by the Van-Arkel-de-Boer process \cite{VanArkel25}.
Validation measurements with the opaque test bench using the new $m=2$ mirrors are planned.

\section{Conclusion}

We demonstrate that the addition of N to the Ti layers results in an effective interdiffusion barrier for multilayer structures of Cu/\TiNx. X-ray diffraction measurements indicate a dilatation of the Ti lattice and a reorientation of the Ti grains from preferential $[002]$ to $[111]$ with increasing nitrogen content. The addition of nitrogen also increases the long-term and temperature stability of the multilayers. X-ray reflectivity measurements show very smooth layer interfaces for layers produced with ion plating sputtering technology.

Neutron reflectivity investigations confirm, that \mbox{$m=2$} Cu/\TiNx non-depolarizing neutron supermirrors with a reflectivity above $\SI{90}{\%}$ can be produced. Compared to Ni/Ti mirrors, the reflectivity at the maximum angle is lower by a few percent only. Background due to neutron absorption in PERC will hence be as low as envisaged in Ref.~\cite{Dubbers08}. The excellent thermal stability will enable baking-out the neutron guide inside the warm bore of the PERC magnet \cite{Wang19} to obtain very good high vacuum conditions. Our new model for roughness growth of the supermirror layers enables an excellent description of the neutron reflectivity measurements.

\section*{Acknowledgment}

This work is in part based upon experiments performed at the TREFF instrument operated by FRM~II and at the X-ray diffractometer of the neutron optics group at the MLZ, and upon experiments at the PF1b instrument of the ILL, and the Q3D instrument at the MLL. The authors gratefully acknowledge the support by T.~Soldner, ILL, with the depolarization measurements and thank A.~Bergmaier for his support with the ERD measurements.

This work was supported by the Priority Program SPP~1491 of the German Research Foundation (DFG), contract No. LA2835/2-1, SCH2708/1-2, and SO1058/1-1.

\bibliography{CuTiSuperMirror}

\end{document}